\documentclass[AER]{AEA}

\usepackage{longtable}
\usepackage{comment}
\usepackage{natbib}
 \usepackage{graphicx}
\usepackage[backref=page]{hyperref}
\hypersetup{
colorlinks = true,
linkcolor = cyan,
citecolor = magenta,
urlcolor = blue,
}

\draftSpacing{1.5}

\begin{document}

\title{Nobel begets Nobel}
\shortTitle{Nobel begets Nobel}
\author{Richard S.J. Tol\thanks{Tol: Department of Economics, University of Sussex, BN1 9SL Falmer, United Kingdom, r.tol@sussex.ac.uk; Institute for Environmental Studies, Vrije Universiteit Amsterdam, The Netherlands; Department of Spatial Economics, Vrije Universiteit Amsterdam, The Netherlands; Tinbergen Institute, Amsterdam, The Netherlands; CESifo, Munich, Germany; Payne Institute of Public Policy, Colorado School of Mines, Golden, CO, USA; College of Business, Abu Dhabi University, Abu Dhabi, United Arab Emirates. Many colleagues kindly answered questions about their ancestry. Bengt Kristr\"{o}m, Peter Dolton and Rocco d'Este had helpful suggestions and comments.}}
\date{\today}
\pubMonth{March}
\pubYear{2023}
\pubVolume{}
\pubIssue{}

\begin{abstract}
I construct the professor-student network for laureates of and candidates for the Nobel Prize in Economics. I study the effect of proximity to previous Nobelists on winning the Nobel Prize. Conditional on being Nobel-worthy, students and grandstudents of Nobel laureates are significantly less likely to win. Professors and fellow students of Nobel Prize winners, however, are significantly more likely to win.
\end{abstract}

\maketitle

\section{Introduction}
A Nobel Prize begets Nobel Prizes, or so the story goes. The departmental, collegial and personal concentration of winners of the Sveriges Riksbank Prize in Economic Sciences in Memory of Alfred Nobel is extraordinary \citep{Tol2022ehjet}. This can be seen as clusters of quality: The best professors come together in the best schools \citep{Ellison2013}, inspire, teach and stimulate each other \citep{Azoulay2010, Borjas2012, BOSQUET2017, Oyer2006}, select the best students \citep{Athey2007}, and train them well \citet{Jones2021}. But it can also be seen through the lens of nepotism \citep{COMBES2008, Hamermesh2003, Laband1994, Medoff2003, Carrell2022, Huber2022}. The Prize Committee solicits nominations from a randomly selected sample of professors of economics\footnote{and all professors in the Nordic countries} \emph{and all living Laureates} \citep{Zuckerman1996}, who may put their proteges forward. \citet{Economist2021} finds that "[t]he best way to win a Nobel is to get nominated by another laureate". That conclusion relies on archival research, which cannot be done for economics as deliberations remain confidential for 50 years. I instead rely on network theory.

\citet{Tol2022ehjet} builds the network of professor-student relations for the Nobel laureates in economics, focussing on formal PhD advisory relations in recent times but including informal mentor-mentee relationships for earlier scholars. There are only four graphs: Pissarides has his own family tree, as do Frisch and Haavelmo, and Allais and Debreu. All other Nobelists are related to one another, sometimes distantly, more often closely. \href{https://academictree.org/econ/tree.php?pid=616505&pnodecount=5\&cnodecount=2\&fontsize=1}{Esther Duflo} is a good example: All three of her professors won the Nobel Prize, as have two of her four grand-professors, one great-grand-professor, one great-great-grand-professor, and one great-great-great-grand-professor. Duflo also illustrates that these close familial ties\footnote{she is also married to a Nobelist} did not stop her from revolutionizing economic methodology.

Besides data on the Nobelists, I also collect data on the candidates for the Nobel Prize\textemdash those economists who have published papers that are highly-cited in economics journals\textemdash and connect them, if possible, to the Nobel family tree. This allows me to test whether well-connected candidates are more likely to win than less-connected ones.

The main contribution of this paper is to show that, conditional on having produced Nobel-worthy research, having a Nobelist as a \emph{student} and \emph{having studied with} a Nobelist significantly and substantially increases the probability of winning the Nobel prize while having a Nobelist as a \emph{professor} reduces the probability of winning. These are associations.

A second contribution is a proposed measure of vertical or cross-distance in a network. In a directed graph, outcloseness measures ancestry and incloseness descent. Cross-closeness measures \emph{shared} ancestry\textemdash in this application, scholars who had the same professor.

A minor contribution is as follows. Statistical inference on a network is difficult because network measures, such as centrality, are descriptive statistics of the population. Changes in a network, on the other hand, can be analyzed statistically using existing methods. The network of Nobelists has changed once a year since 1970. It could have changed in many different ways but it has changed in one particular way. That is, changes in a network can be analyzed using standard selection models.

The paper proceeds as follows. Section \ref{sc:data} discusses the data and methods used. Section \ref{sc:results} presents the results. Section \ref{sc:conclude} concludes.

\section{Data and methods}
\label{sc:data}

\subsection{Data}
The Nobel network documents, for the most part, the relationships between PhD advisers and candidates. However, PhDs are not standardized today and variation was greater in the past. Roy Harrod, for instance, had studied history before being privately mentored in economics by John Maynard Keynes. The network therefore also includes more general mentor-mentee or professor-student relations. Data were taken from \href{https://academictree.org/econ/}{AcademicTree}, complemented by data from \href{https://genealogy.repec.org/}{RePEc}, \href{https://www.genealogy.math.ndsu.nodak.edu/}{Mathematics Genealogy}, Wikipedia, curriculum vitae, PhD theses, and direct inquiries to the people involved. Data are stored on \href{https://academictree.org/econ/}{AcademicTree}. The network in \citet{Tol2022ehjet} ends in 2021, is here updated with Bernanke, Diamnond and Dybvig. Up to 15 generations are included. Christian Haussen, Christian Heyne, August Schlegel and Pierre Varignon are the common ancestors who connect 82 of the 87 Nobelists. These are not household names, indeed not economists. None of the Classical economists we find in textbooks appear in the network, and only two of the renowned neo-Classicists (Menger and Marshall, the latter, ironically, via Keynes). \citet{Tol2022ehjet} finds that Karl Knies, who taught John Bates Clark, Eugen B\"{o}hn von Bawerk, Richard Ely, and Edwin Seligman, among others, is the central-most professor, followed by Wassily Leontief, the professor of Paul Samuelson, Thomas Schelling, Vernon Smith, Robert Solow, and others. Knies' central role is perhaps surprising. He was a member of the Historical School, arguing that economics should be an empirical science just as it turned to theory.\footnote{Knies would have delighted in the empirical work of Angrist, Card and Duflo, who are his academic descendants.} But while the intellectual foundations of economics lie in Great Britain, the roots for training research economists lie in Germany. Young Americans aspiring to be economists saw Knies as the man to help them meet that ambition and they passed the lessons learned to the next generation.

Nobel laureates are readily identified. Nobel candidates are not. There is much speculation about what it takes to win. A necessary condition is to have shaped or created a substantial field of economics; to have opened a new line of inquiry, either thematically or methodologically. This is operationalized by citations in the economics literature, which in recent years typically measure in the tens of thousands, concentrated on a few seminal papers \citep[e.g.,][]{Mixon2017, Chan2018}. Clarivate's \href{https://clarivate.com/hall-of-citation-laureates/}{Citation Laureates} meet these criteria and indeed many Citation Laureates later won the Nobel prize\textemdash although they missed two of the three Laureates of 2022.

Clarivate's list is arguably incomplete. Cross-checking with the IDEAS/RePEc \href{https://ideas.repec.org/top/top.item.nbcites.html}{list of most cited papers}, I added Tim Bollerslev. Cross-checking with the IDEAS/RePEc \href{https://ideas.repec.org/top/top.person.wsccites.html}{list of highly cited authors}, I added Andrei Shleifer, John Campbell and Robert Vishny.\footnote{I did not use the rankings of \href{https://research.com/scientists-rankings/economics-and-finance}{Research.com} and \href{https://scholar.google.com/citations?view_op=search_authors&hl=en&mauthors=label:economics}{Google Scholar}, because both platforms have issues with citation counts and identification of scholars.} These candidates are based on clear criteria that can be applied to data available for recent times.

I added Alvin Hansen, Harold Hotelling, Frank Knight, Abba Lerner, Ludwig von Mises and Oskar Morgenstern, who all died too soon to make it onto any recent lists but would have been worthy. I added Sanford Grossman as a John Bates Clark medalist who saw the co-authors of his most-cited papers win the Nobel prize for something else.\footnote{There is an unwritten convention that economists can win only one Nobel prize. John Bardeen won the Nobel Prize in physics twice, Frederick Singer and Barry Sharpless won twice in chemistry, and Marie Sk\l odowska Curie won physics and chemistry.} Fischer Black would have shared Myron Scholes's Nobel Prize, and Jean-Jacques Laffont Jean Tirole's had they lived long enough. Although David Kreps is a Clarivate Citation Laureate, his co-author Evan Porteus is not. Michael Jensen is a Citation Laureate, but his co-author William Meckling passed too soon for that honor. I further added Guillermo Calvo, Lionel McKenzie, Jacob Mincer, and Henri Theil for their work on sticky prices,\footnote{and for having a fairy named after him} general equilibrium, labour, and two-stage least squares, respectively, and Roy Harrod, Kelvin Lancaster, and Mancur Olson for their research on economic growth, the second best, and public goods, respectively. I also added Francine Blau, Ester Boserup, Edith Penrose, and Joan Robinson for their work on inequality, development, firms, and capital, respectively. The full list of candidates and Nobelists is given in Table \ref{tab:names} in the Appendix. I conduct a sensitivity analysis on the list of candidates below.

As with the Nobel laureates, I collected information about their ancestry from the \href{https://academictree.org/econ/}{Academic Tree}. If there was no entry, I checked \href{https://genealogy.repec.org/}{RePEc Genealogy}, \href{https://www.mathgenealogy.org/index.php}{Mathematics Genealogy}, Wikipedia, CVs, and published theses. If all that failed, I wrote to the candidate or a close associate. I added the information thus collected to the \href{https://academictree.org/econ/}{Academic Tree}. The data were transferred to Matlab for visualization and analysis. Code and data are available on \href{https://github.com/rtol/NobelNetwork}{GitHub}.

The final network has 668 nodes and 808 edges. This is too large to meaningfully display in a graph \citep[see][for an attempt]{Tol2022ehjet}. Figures \ref{fig:leontief}-\ref{fig:modigliani} display subgraphs for the descendants of four Nobel laureates, Leontief, Arrow, Tinbergen and Modigliani. The Leontief graph has four Nobelists in the first generation, eight in the second generation, two in the third generation, and one in the fifth. The graph also shows a number of plausible candidates.

For every Laureate and candidate, I collected year of birth, year of death (if applicable), the year of winning, gender, \textit{alma mater}, and one-digit JEL classifier using Wikipedia as the main source of information.\footnote{Recall that the first Nobel prizes were awarded in 1969, for research done decades before that and published without JEL codes.} The long list of candidates and Laureates is turned into a short-list of candidates for each year when they (\textit{i}) are alive, (\textit{ii}) are over 40, and (\textit{iii}) have not yet won. This then implies a zero-one variable for people who could have won (0) and people who did (1) for every year from 1970 to 2021. 1969 is excluded because no one had any connection to a previous Nobelist.\footnote{Jan Tinbergen was the student and grandstudent of two prominent physicists, Ehrenfest and Boltzmann, who did not win the Nobel prize, however. Tjalling Koopmans, on the other hand, has three Nobel laureates in physics (Bohr, Thomson, Strutt) and one in chemistry (Rutherford) in his ancestry and, of course, one in economics (Tinbergen). Daniel Kahneman is a distant descendant from a medicine laureate (Sherrington).}

The variable of interest is the proximity (defined below) of the candidates in year $t$ to the Nobelists of years $s<t$. I distinguish between the proximity to academic ancestors, descendants, and peers. For ancestors and peer, I further distinguish between recent and earlier laureates and, for ancestors, between living and dead professors.

\subsection{Methods}
The network of professor-student relationships can be represented by a graph, more specifically, a directed acyclic graph or polytree. The distance from a node $i$ in a graph to the rest of this graph can be measured by the H\"{o}lder mean
\begin{equation}
\label{eq:holder}
    D_{i,t}(h) = \left (\frac{1}{N_t} \sum_{j \in Nobel} D_{j,i,t}^h \right )^\frac{1}{h}
\end{equation}
where $D_{j,i}$ is the distance from node $i$ to any node $j$, that is, the number of edges between the $i$ and $j$. As the interest is in \emph{Nobel} ancestry, attention is restricted to the distance to Nobelists. $N_t$ is therefore the number of previous winners of the Nobel Prize at time $t$.

It is common to set $h=1$. The H\"{o}lder mean is then the familiar arithmetic mean. However, $D_{i,t}(1) = \infty$ unless scholar $i$ descends from \emph{all} previous Nobelists. There is no such scholar.

For $h=-1$, the H\"{o}lder mean is the harmonic mean, which is bounded if some nodes in the network cannot be reached. In other words, the harmonic mean applies to connected as well as unconnected subgraphs: For unreachable nodes $D_{j,i} = \infty$ so $1/D_{j,i} = 0$. \citet{Marchiori2000} propose this as a measure of distance.

For ease of interpretation, I follow \citet{GilSchmidt1996}, who propose the \emph{inverse} of the harmonic mean as a measure of closeness $C_{i,t}(p) = D_{i,t}(h)^{-1}$. Scholars who have no Nobelists in their ancestry score 0; the score increases with more and more proximate Nobel laureates. This is an \emph{outcloseness} measure. Outcloseness on a polytree measures ancestry. According to this measure, two students of the same Nobelist are both close to their professor, but not to each other (see below).

Recall that I do not use the proximity to all nodes, but only to the Nobel ones. In Equation (\ref{eq:holder}), $N$ is the number of Nobelists and $j$ sums over them. Concretely, therefore, a candidate receives one point for every professor who won the Nobel Prize, half a point for every grandprofessor who did, a third of a point for every Nobel great-grandprofessor, and so on, and zero points for academic ancestors who are not laureates. Note that many economists in the network have more than one Nobel ancestor. The total number of points is then divided by the total number of laureates \emph{at the time}.

The harmonic mean distance emphasizes proximity at the expense of distal relationships: Consider a student with one Nobel professor (distance 1) and one Nobel great-grandprofessor (distance 3). The arithmetic mean distance is 2, the harmonic mean distance is 1.5. That is, the harmonic mean is skewed towards closer relationships.

I also compute an \emph{incloseness} measure, replacing $D_{j,i}$ by $D_{i,j}$ in Equation (\ref{eq:holder}). This measures the distance to Nobel \emph{students}.

Besides outcloseness (ancestry) and incloseness (descent), which are based on the vertical distance \emph{between} generations, I also measure crosscloseness, which is based on the horizontal distances \emph{within} generations. Define the proximity of degree 1 (2) between nodes $i$ and $j$ as one minus the number of shared (grand) professors divided by the total number of (grand) professors of $i$ and $j$. This measure is one for full siblings (double first cousins).\footnote{Proximity of degree $n$ is readily defined but the language becomes tedious. Horizontal proximity based on shared descent can be defined in an analogous way but makes little sense in the current context.} The crosscloseness of node $i$ is then the inverse of the harmonic mean horizontal proximity to all other nodes $j$. See \citet{Tol2023nobelware} for further discussion of distance measures and the required software.

As the number of Nobel laureates grows over time, proximity to Nobelists is a non-stationary measure. I therefore scale $C_{i,t}$ by $\max_i C_{i,t}$. Proximity is thus replaced by \emph{relative} proximity, where the closest candidate in any year scores one and all others score less than that.

\section{Results}
\label{sc:results}
Table \ref{tab:results} shows the results of eight regressions. The dependent variable is zero-one, so I use logit and probit. I estimate the model with and without year fixed effects, with and without fixed effects for the \textit{alma mater}, and with and without field fixed effects. I treat all previous Nobelists equally (see below). In all eight specifications, the proximity to Nobel students is significant and positive. That is, conditional on being a candidate for the Nobel prize, the probability of winning increases if your students have won before you. This pattern started early: Leontief won after his student Samuelson, Hayek after his student Hicks. It continues today. Wilson won after his students Holmstr{\"o}m and Roth (and together with a third student, Milgrom.) Angrist (Card) won after his (grand)student Duflo. One possible explanation is that the surge of interest that accompanies a Nobel Prize leads to a re-appreciation of the foundations on which that work was built.

Proximity to Nobel peers is also positive and significant in all specifications. You are more likely to win if someone from the same PhD programme has won before you. As noted above, three of Wilson's students won the Nobel prize. Arrow, Solow and Samuelson also have three Nobel students, Leontief has four. It is these concentrations of academic siblings and, for Leontief, cousins that explain the significance of the coefficient for peer proximity.

\begin{figure}
    \centering
    \includegraphics[width=\textwidth]{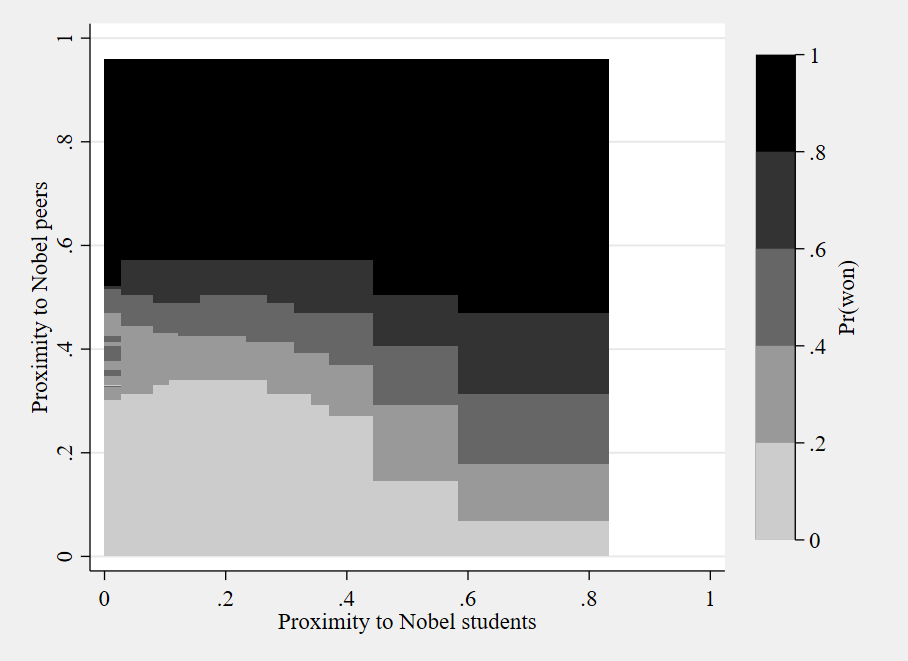}
    \caption{Probability, conditional on being a candidate, of winning the Nobel prize in economics as a function of the relative proximity to a previously enNobeled student and peer.}
    \label{fig:prob}
\begin{figurenotes}
Predicted probability according to the logit model with year fixed effects. See Table \ref{tab:results}.
\end{figurenotes}
\end{figure}

Table \ref{tab:results} shows the estimated coefficients. Figure \ref{fig:prob} shows the effect size, that is, the predicted probability of winning the Nobel prize, using the logit model with year fixed effects, as a function of the relative proximity to Nobel students and peers. The effect size is substantial. In the bottom left corner, those candidates without Nobel students and peers, have a predicted annual win probability of 20\% or less. This probability is over 80\% for those who are closest to Nobel descendants and siblings.

Proximity to Nobel professors is significant but negative: Students of Nobel laureates are less likely to win. However, there is a strong positive correlation between proximity to Nobel professors and Nobel peers. If the latter is dropped from the regression, the effect of Nobel professors becomes positive but insignificant. Recall, furthermore, that all estimates are conditional on being a candidate. Most candidates have some connection to a previous Laureate. Successful candidates tend to have connections to both Nobel professors and Nobel peers\textemdash which is why, conditional on being a candidate and controlling for peers, proximity to Nobel professors reduces the probability of winning the Nobel prize.

The inclusion of fixed effects for the \textit{alma mater} does not change the results in a meaningful way. Observations are dropped because some universities have candidates but no winners. Cambridge, Chicago, Harvard, LSE, MIT, Oxford, Princeton, Stanford and Yale have significant negative effects for both probit and logit, Carnegie Mellon for logit only. Leiden and Leningrad have positive effects for logit only. Conditional on being a candidate, getting a PhD from a smaller, less prestigious programme increases the probability of winning the Nobel prize.

Field fixed effects again leave the main results unaffected. Dummies are significant for JEL-codes D (micro), E (macro), K (law) and Q (environment). Researchers in this field are significantly less likely to win the Nobel Prize, conditional on being a candidate.

Table \ref{tab:results} also includes a gender dummy. Female candidates are less likely to win than male candidates, but this effect is insignificant. Although there are claims to the contrary \citep[but see][]{Liang2020}, women are not discriminated against in this regard.

Table \ref{tab:split} sheds some light on the mechanism. One hypothesis is that Nobel prizes are won on the recommendation of Laureates. Dead people do not write nomination letters, so I split the sample into deceased and living professors. However, no one has won the Nobel Prize after a student or peer has won and died. You can further hypothesize that recent Nobelists more eagerly exercise their powers of patronage. So I split the sample into professors and peers who won within the last 10 years and earlier winners. Professors win within 10 years of their Nobel students.

Table \ref{tab:split} shows that proximity to Nobel professors remains negative when I distinguish between living and deceased professors. The effect size is smaller and insignificant for \emph{deceased} professors. (Don't get any ideas, guys!) It's only weakly significant for living professors. The same result is found when I split proximity to Nobel professors between those who won in the last decade and those who won more than 10 years ago. In the right-most columns, I further split Nobel peers into recent and earlier winners. Earlier laureates have no significant impact, but recent ones do\textemdash the effect size increases. In this specification, recently enNobeled professors also have a highly significant, negative effect.

These results suggest that nominations by previous Nobelists play a role. To strengthen the argument, I collected the citation histories of all Nobelists from the Web of Science. I regressed the annual citation number on a linear time trend, and a dummy for the years after an academic relative won the Nobel Prize. See \citet{Bjork2014, Sangwal2015, Zhang2019swan, Frey2020, Kosmulski2020} for more advanced analysis. The t-statistics for these dummies are shown in Figures \ref{fig:prof}, \ref{fig:stud} and \ref{fig:peer}. The awarding of the Nobel Prize to a professor had a statistically significant impact on citations to his students in 13 cases (who later went on to win anyway), a positive effect in 3 cases, and no significant effect in 7. See Figure \ref{fig:prof}.

The Nobel Prize going to a student positively (negatively) affected the citations of their professors in two cases (one case), and had no statistically significant effects in 5 cases. See Figure \ref{fig:stud}. The Nobel Prize positively (negatively) affects the citation record of fellow students in 6 (12) cases, and not at all in 14 cases. See Figure \ref{fig:peer}. Putting all these results together, previous Nobel laureates do not affect their relatives' chances of winning because they draw more citations to their work.



Table \ref{tab:who} restricts the number of candidates, first by excluding the candidates identified by me and denoted as ``ad hoc'' in Table \ref{tab:names}, and then by excluding those as well as the candidates found at IDEAS/RePEc. With fewer candidates, the average probability of winning goes up. It does not materially affect the results. Coefficients and significance are almost the same in the limited samples as in the full sample.


I did not include more control variables. Including a quality indicator, such as the number or concentration of citations, would just confirm that all candidates are Nobel-worthy\textemdash the Nobel prize is not handed out mechanically. Designing a quality indicator that is robust over five decades and across subdisciplines is not easy and not attempted here. Figure \ref{fig:cumcit} shows the total number of citations at the time the Nobel Prize was awarded. There is a general upward trend as the discipline has grown. There is wide variation around that trend. Arthur Lewis had one citation\textemdash he wrote books, which are not counted in the \textit{Web of Science}, while Eugene Fama had 34,778\textemdash the words ``at last'' were probably the most common in response to the announcement in 2013.

Some commentators discern a pattern through which different parts of the profession get awarded on the basis of a pre-determined rota. There is undoubtedly some of that going on. Last year's runner-up would be this year's favourite. Subfields or schools that feel overlooked may be more eager to submit nominations.\footnote{If that is what they do. I used my first spell as a nominator to argue for Thomas Schelling, my second spell to argue for Anne Krueger. The Nobel laureates closest to my own research are William Nordhaus and Robert Wilson.} Pigeon-holing candidates is subjective and difficult, particularly since Nobelists tend to win for having broken the mold. Discerning the preferences of the members of selection committee is harder still, let alone the dynamics of the discussions within the committee, the composition of which changes over time. Documenting the sympathies and antipathies that increase and decrease the chance of winning is almost impossible.\footnote{For instance, it may seem peculiar that there is a Nobel prize for discrete choice but not for two-stage least squares, a method that is used more widely. It is not peculiar for those who know.}

\section{Discussion and conclusion}
\label{sc:conclude}
I test whether academic relations of previous laureates are more likely to win the Nobel memorial prize in economics. Conditional on being a candidate, the \emph{professors} and \emph{peers} of Nobelists are more likely to win but their \emph{students} are less likely to win. The positive impact of peers is limited to fellow students who won the Nobel prize less than 11 years ago. The negative impact of professors only holds for those who won less than 11 years ago or are still alive. There is no consistent impact of winning the Nobel prize on the citations to the papers of Laureates' students, peers, or professors. Nobelists lobbying for their professors and peers can therefore not be excluded. Your best bet to win a Nobel prize is to make sure your (fellow) students win one first.

There are three big gaps in this research. A study of the archives of the Nobel committee would shed more light on nominations, discussions, and group dynamics. Unfortunately, most of these archives are sealed. It will take a few more decades before a sufficiently large sample is available for study. The second gap is that the network used is the network of \emph{formal} advisory relationships. \emph{Informal} mentoring is just as important but hard to document for people who did not leave an autobiography, extensive correspondence, or in-depth interviews. The co-authorship network may be important too\textemdash data were not collected for this paper\textemdash after careful controlling for changes over time in publication norms and cultures and the size of the economics profession.

The third, and arguably most important gap is the candidacy. The results above are all conditional on having established a track record that is Nobel-worthy. The current paper is silent on the question how to become Nobel-worthy\textemdash see \citet{Korom2021} for an attempt. In future research, a long list could be created of everyone who has published in a Top 5 journal or every economist with a paper with over 1,000 citations\textemdash perhaps supplemented with semantic analysis \citep{Iaria2018} or citation analysis \citep{Uzzi2013} to capture novelty\textemdash and use that to test what it takes to progress from not-impossible to a plausible candidate.

It is an open question how the networks of Nobel candidates differ from other economists\textemdash particularly to what extent excellent people group together, in departments and co-author networks, and how such social dynamics propel researchers to new heights. These issues are postponed to future research.

\begin{table}[htbp]\centering\scriptsize
\def\sym#1{\ifmmode^{#1}\else\(^{#1}\)\fi}
\caption{Probability of winning the Nobel Prize.\label{tab:results}}
\begin{tabular}{l*{8}{c}}
\hline\hline
                    &\multicolumn{1}{c}{Logit}&\multicolumn{1}{c}{Probit}&\multicolumn{1}{c}{Logit}&\multicolumn{1}{c}{Probit}&\multicolumn{1}{c}{Logit}&\multicolumn{1}{c}{Probit}&\multicolumn{1}{c}{Logit}&\multicolumn{1}{c}{Probit}\\
\hline
female              &      -0.196         &      0.0263         &      -0.238         &      0.0163         &      -1.633         &      -0.308         &      -0.523         &      0.045         \\
                    &     (-0.27)         &      (0.10)         &     (-0.32)         &      (0.06)         &     (-1.63)         &     (-0.79)         &     (-0.63)         &     (0.16)       \\
[1em]
Proximity to\\
Nobel profs             &      -2.076\sym{***}&      -0.771\sym{**} &      -1.907\sym{**} &      -0.799\sym{**} &      -2.254\sym{*}  &      -1.082\sym{**} &      -2.152\sym{**} &      -0.983\sym{***} \\
                    &     (-3.43)         &     (-3.12)         &     (-3.00)         &     (-3.01)         &     (-2.50)         &     (-2.91)         &     (-2.95)         &     (-3.26)     \\
[1em]
Nobel students             &       3.642\sym{***}&       1.883\sym{***}&       4.259\sym{***}&       2.201\sym{***}&       6.957\sym{***}&       3.494\sym{***}&       5.049\sym{***}&       2.675\sym{***}\\
                    &      (4.15)         &      (3.83)         &      (3.32)         &      (3.44)         &      (3.51)         &      (3.47)         &      (3.36)         &      (3.70)         \\
[1em]
Nobel peers           &       10.93\sym{***}&       5.228\sym{***}&       11.59\sym{***}&       5.522\sym{***}&       16.11\sym{***}&       7.212\sym{***}&       14.22\sym{***}&       6.62\sym{***}\\
                    &     (13.85)         &     (13.83)         &     (13.46)         &     (13.65)         &     (12.54)         &     (13.21)         &     (12.58)         &     (13.11)         \\
\hline
Year fixed effects & No & No & Yes & Yes & Yes & Yes & Yes & Yes\\
\textit{Alma mater} fixed effects & No & No & No & No & Yes & Yes & No & No\\
Field fixed effects & No & No & No & No & No & No & Yes & Yes\\
Observations        &        4899         &        4899         &        4899         &        4899         &        4245         &        4245         &        4851         &        4851         \\
\hline\hline
\multicolumn{9}{l}{\footnotesize \textit{t} statistics in parentheses}\\
\multicolumn{9}{l}{\footnotesize \sym{*} \(p<0.05\), \sym{**} \(p<0.01\), \sym{***} \(p<0.001\)}\\
\end{tabular}
\end{table}

\begin{table}[htbp]\centering\scriptsize
\def\sym#1{\ifmmode^{#1}\else\(^{#1}\)\fi}
\caption{Probability of winning the Nobel Prize.\label{tab:split}}
\begin{tabular}{l*{8}{c}}
\hline\hline
                    &\multicolumn{1}{c}{Logit}&\multicolumn{1}{c}{Probit}&\multicolumn{1}{c}{Logit}&\multicolumn{1}{c}{Probit}&\multicolumn{1}{c}{Logit}&\multicolumn{1}{c}{Probit}&\multicolumn{1}{c}{Logit}&\multicolumn{1}{c}{Probit}\\
\hline
female              &      -0.238         &      0.0163         &      -0.221         &      0.0212         &      -0.221         &      0.0212   & 0.182  & 0.154    \\
                    &     (-0.32)         &      (0.06)         &     (-0.30)         &      (0.08)         &     (-0.30)         &      (0.08)     & (0.24)  & (0.55)  \\
[1em]
Proximity to & & & & & & \\
 Nobel students             &       4.259\sym{***}&       2.201\sym{***}&       4.218\sym{***}&       2.179\sym{***}&       4.218\sym{***}&       2.179\sym{***} & 4.990\sym{***} & 2.153\sym{***}\\
                    &      (3.32)         &      (3.44)         &      (3.31)         &      (3.42)         &      (3.31)         &      (3.42) & (3.13) & (3.26)       \\
[1em]
 Nobel peers           &       11.59\sym{***}&       5.522\sym{***}&       11.43\sym{***}&       5.448\sym{***}&       11.43\sym{***}&       5.448\sym{***} & & \\
                    &     (13.46)         &     (13.65)         &     (13.40)         &     (13.64)         &     (13.40)         &     (13.64)     & &    \\               
[1em]
 earlier enNobeled peers & & & & & & & -0.116 & 0.341\\
  & & & & & & & (-0.08) & (0.50)\\
[1em]
recently enNobeled peers & & & & & & & 18.28\sym{***} & 8.869\sym{***}\\
  & & & & & & & (7.96) & (8.16)\\
[1em]
 Nobel profs              &      -1.907\sym{**} &      -0.799\sym{**} &                     &    & &                 &                     &                     \\
                    &     (-3.00)         &     (-3.01)         &                     &                     &                     &   & &                  \\
[1em]
 deceased Nobel profs          &                     &                     &      -0.911         &      -0.413         &                     &                     & & \\
                    &                     &                     &     (-1.18)         &     (-1.35)         &                     &  & &                    \\
[1em]
 living Nobel profs          &                     &                     &      -1.428\sym{*}  &      -0.582\sym{*}  &                     &           & &          \\
                    &                     &                     &     (-2.49)         &     (-2.43)         &                     &    & &                 \\
[1em]
 earlier enNobeled profs       &                     &                     &                     &                     &      -0.911         &      -0.413 & -1.208   & -0.450     \\
                    &                     &                     &                     &                     &     (-1.18)         &     (-1.35)  & (-1.31)      & (-1.28) \\
[1em]
 recently enNobeled profs        &                     &                     &                     &                     &      -1.428\sym{*}  &      -0.582\sym{*}  &  -2.024\sym{***} & -0.892\sym{***}\\
                    &                     &                     &                     &                     &     (-2.49)         &     (-2.43)  & (-3.10)   & (-3.15)    \\
\hline
Year fixed effects & Yes & Yes & Yes & Yes & Yes & Yes & Yes & Yes\\
\textit{Alma mater} fixed effects & No & No & No & No & No & No & No & No\\
Field fixed effects & No & No & No & No & No & No & No & No\\
Observations        &        4899         &        4899         &        4899         &        4899         &        4899         &        4899    & 4899 & 4899     \\
\hline\hline
\multicolumn{7}{l}{\footnotesize \textit{t} statistics in parentheses}\\
\multicolumn{7}{l}{\footnotesize \sym{*} \(p<0.05\), \sym{**} \(p<0.01\), \sym{***} \(p<0.001\)}\\
\end{tabular}
\end{table}

\begin{table}[htbp]\centering\scriptsize
\def\sym#1{\ifmmode^{#1}\else\(^{#1}\)\fi}
\caption{Probability of winning the Nobel Prize.\label{tab:who}}
\begin{tabular}{l*{6}{c}}
\hline\hline
&\multicolumn{2}{c}{all}&\multicolumn{2}{c}{without \textit{ad hoc}} &\multicolumn{2}{c}{without IDEAS/RePEc} \\
                    &\multicolumn{1}{c}{Logit}&\multicolumn{1}{c}{Probit}&\multicolumn{1}{c}{Logit}&\multicolumn{1}{c}{Probit}&\multicolumn{1}{c}{Logit}&\multicolumn{1}{c}{Probit}\\
\hline
female              &      -0.238         &      0.0163         &       0.306         &       0.218         &       0.296         &       0.213         \\
                    &     (-0.32)         &      (0.06)         &      (0.42)         &      (0.79)         &      (0.40)         &      (0.77)         \\
[1em]
Proximity to \\
Nobel profs              &      -1.907\sym{**} &      -0.799\sym{**} &      -1.992\sym{**} &      -0.860\sym{**} &      -1.965\sym{**} &      -0.848\sym{**} \\
                    &     (-3.00)         &     (-3.01)         &     (-3.14)         &     (-3.21)         &     (-3.10)         &     (-3.16)         \\
[1em]
Nobel students             &       4.259\sym{***}&       2.201\sym{***}&       4.091\sym{**} &       2.136\sym{***}&       4.076\sym{**} &       2.128\sym{**} \\
                    &      (3.32)         &      (3.44)         &      (3.16)         &      (3.29)         &      (3.15)         &      (3.28)         \\
[1em]
Nobel peers           &       11.59\sym{***}&       5.522\sym{***}&       11.50\sym{***}&       5.529\sym{***}&       11.45\sym{***}&       5.505\sym{***}\\
                    &     (13.46)         &     (13.65)         &     (13.31)         &     (13.48)         &     (13.21)         &     (13.38)         \\
\hline
Year fixed effects & Yes & Yes & Yes & Yes & Yes & Yes\\
\textit{Alma mater} fixed effects & No & No & No & No & No & No\\
Field fixed effects & No & No & No & No & No & No\\
Observations        &        4899         &        4899         &        4432         &        4432         &        4340         &        4340         \\
\hline\hline
\multicolumn{7}{l}{\footnotesize \textit{t} statistics in parentheses}\\
\multicolumn{7}{l}{\footnotesize \sym{*} \(p<0.05\), \sym{**} \(p<0.01\), \sym{***} \(p<0.001\)}\\
\end{tabular}
\end{table}

\bibliographystyle{aea}
\bibliography{master}

\appendix
\setcounter{page}{1}
\renewcommand{\thepage}{A\arabic{page}}
\setcounter{equation}{0}
\renewcommand{\theequation}{A\arabic{equation}}
\setcounter{table}{0}
\renewcommand{\thetable}{A\arabic{table}}
\setcounter{figure}{0}
\renewcommand{\thefigure}{A\arabic{figure}}

\section{Family relationships}
\label{sc:families}

\subsection{Nobel professors}
Nobelists who won after their Nobel professors:
\begin{itemize}
    \item Koopmans (1975) after Tinbergen (1969)
    \item Friedman (1976) after Kuznets (1971)
    \item Klein (1980) after Samuelson (1970)
    \item Solow (1987) after Leontief (1973)
    \item Haavelmo (1989) after Frisch (1969)
    \item Markowitz (1990) after Friedman (1976)
    \item Fogel (1993) after Kuznets (1971)
    \item Harsanyi (1994) after Arrow (1972)
    \item Mirrlees (1996) after Stone (1984)
    \item Merton (1997) after Samuelson (1970)
    \item Scholes (1997) after Miller (1990)
    \item Akerlof (2001) after Solow (1987)
    \item Smith (2002) after Leontief (1973)
    \item Schelling (2005) after Leontief (1973)
    \item Phelps (2006) after Tobin (1981)
    \item Maskin (2007) after Arrow (1972)
    \item Myerson (2007) after Arrow (1972)
    \item Williamson (2009) after Simon (1978)
    \item Diamond (2010) after Solow (1987)
    \item Fama (2013) after Miller (1990)
    \item Hansen (2013) after Sims (2011)
    \item Shiller (2013) after Modigliani (1985)
    \item Tirole (2014) after Maskin (2007)
    \item Deaton (2015) after Stone (1984)
    \item Nordhaus (2018) after Solow (1987)
    \item Romer (2018) and Lucas (1995)
    \item Banerjee (2019) after Maskin (2007)
\end{itemize}

\subsection{Nobel students}
Nobelists who won after their Nobel students:
\begin{itemize}
    \item Allais (1988) after Debreu (1983)
    \item Leontief (1973) after Samuelson (1970)
    \item Schelling (2005) after Spence (2001)
    \item Hurwicz (2007) after McFadden (2000)
    \item Fama (2013) after Scholes (1997)
    \item Wilson (2020) after Roth (2012) and Holmstrom (2016)
    \item Angrist (2021) after Duflo (2019)
\end{itemize}

\subsection{Nobel peers}
Nobelists who won after their Nobel peers:
\begin{itemize}
    \item Ohlin (1977) after Myrdal (1974)
    \item Simon (1978) after Friedman (1976)
    \item Tobin (1981) after Samuelson (1970)
    \item Stone (1984) after Meade (1977)
    \item Buchanan (1986) after Stigler (1982)
    \item Solow (1987) after Samuelson (1970)
    \item Markowitz (1990) after Modigliani (1985)
    \item Coase (1991) after Stigler (1982) and Buchanan (1986)
    \item Fogel (1993) after Friedman (1976)
    \item Lucas (1995) after Becker (1992)
    \item Merton (1997) after Klein (1980)
    \item Stiglitz (2001) after Sen (1998)
    \item Smith (2002) after Samuelson (1970) and Solow (1987) 
    \item Schelling (2005) after Samuelson (1970), Solow (1987) and Smith (2002)
    \item Hurwicz (2007) after Modigliani (1985) and Markowitz (1990)
    \item Eric Maskin (2007) after Harsanyi (1994)
    \item Roger Myerson (2007) after Harsanyi (1994)
    \item Diamond (2010) after Akerlof (2001)
    \item Mortensen (2010) after Prescott (2004)
    \item Shapley (2012) after Nash (1994)
    \item Deaton (2015) after Mirrlees (1996)
    \item Holmstrom (2016) after Roth (2012)
    \item Nordhaus (2018) after Akerlof (2001) and Diamond (2010)
    \item Banerjee (2019) after Tirole (2014)
    \item Milgrom (2020) after Roth (2012) and Holmstrom (2016)
\end{itemize}

\section{Additional material}

\begin{figure}
    \centering
    \includegraphics[width=\textwidth]{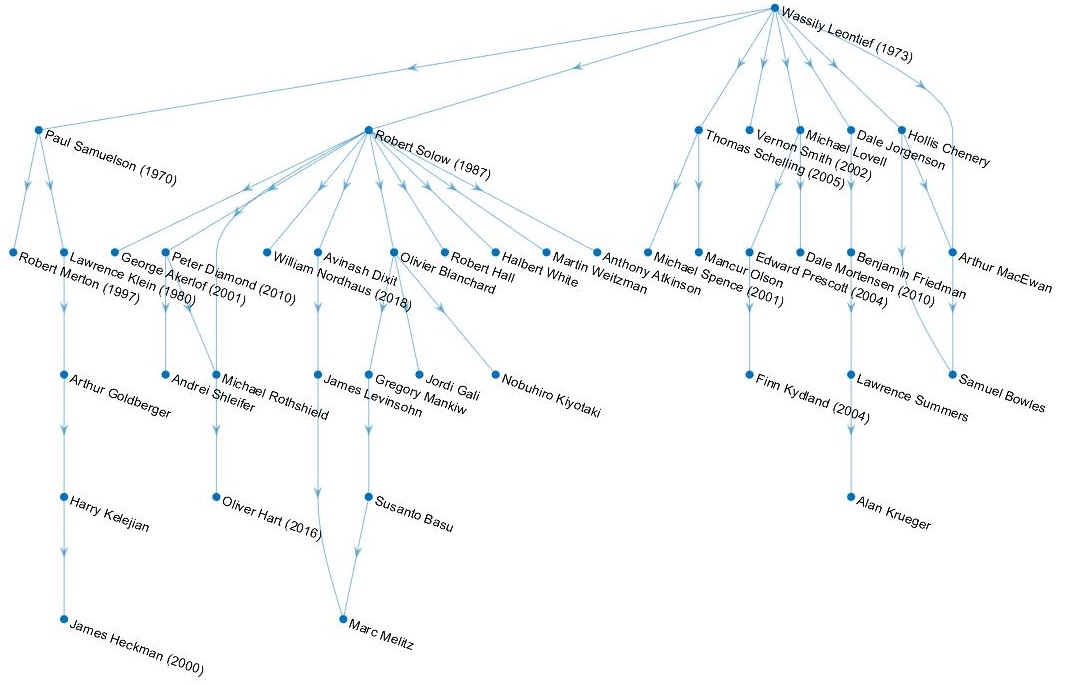}
    \caption{Academic descendants of Leontief.}
    \label{fig:leontief}
\end{figure}

\begin{figure}
    \centering
    \includegraphics[width=\textwidth]{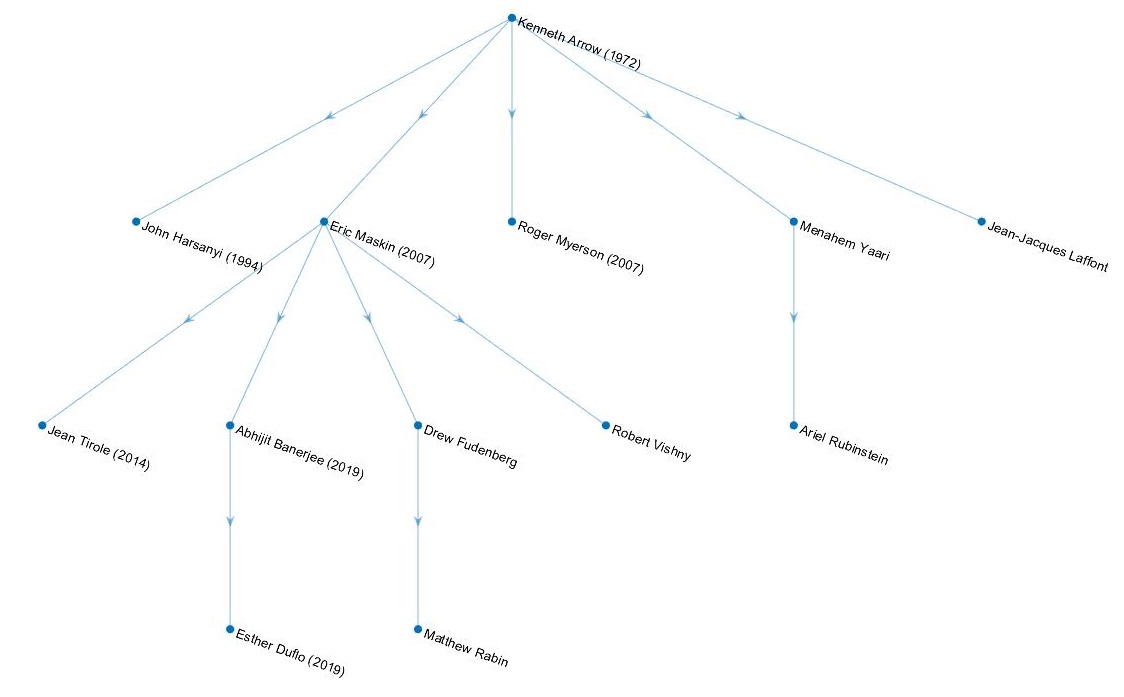}
    \caption{Academic descendants of Arrow.}
    \label{fig:arrow}
\end{figure}

\begin{figure}
    \centering
    \includegraphics[width=\textwidth]{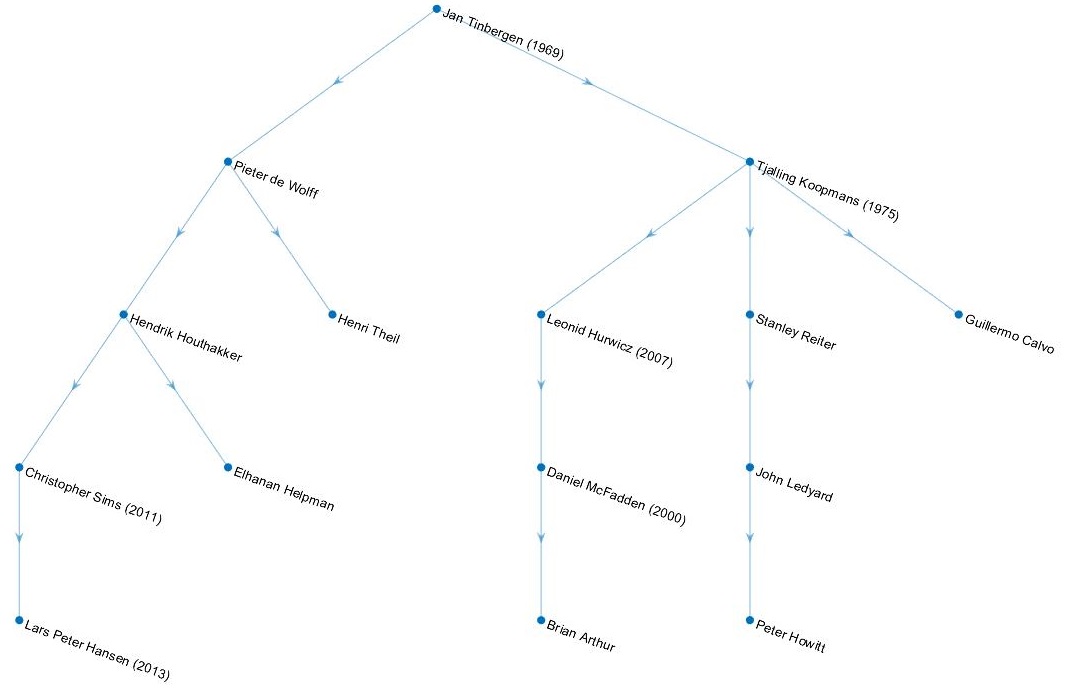}
    \caption{Academic descendants of Tinbergen.}
    \label{fig:tinbergen}
\end{figure}

\begin{figure}
    \centering
    \includegraphics[width=\textwidth]{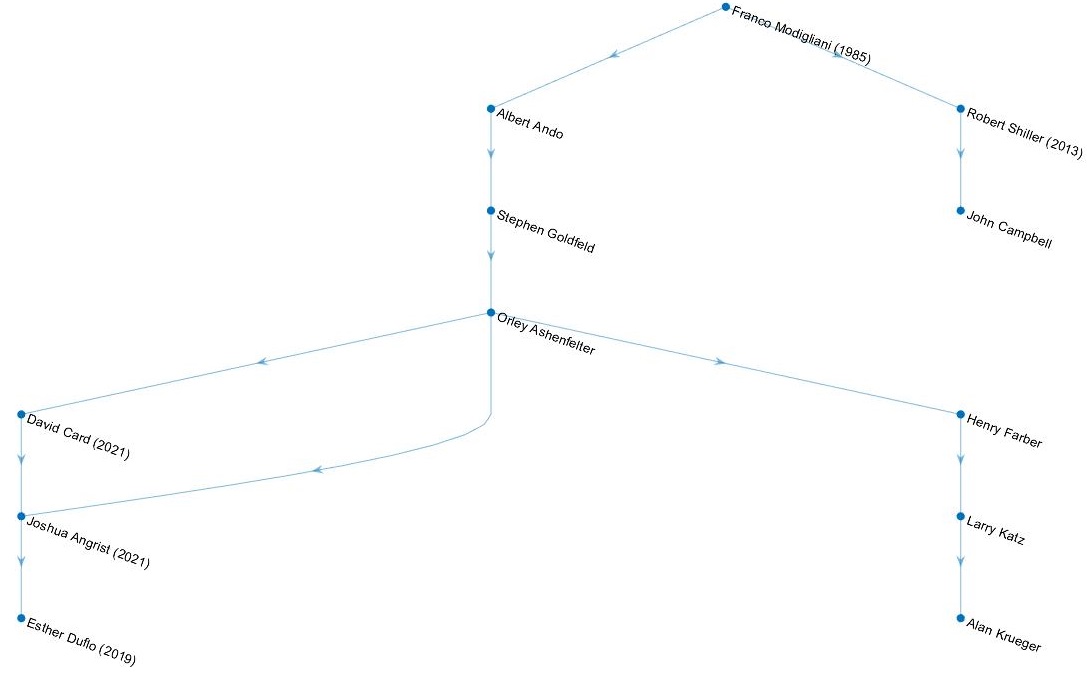}
    \caption{Academic descendants of Modigliani.}
    \label{fig:modigliani}
\end{figure}

\begin{figure}
    \centering
    \includegraphics[width=\textwidth]{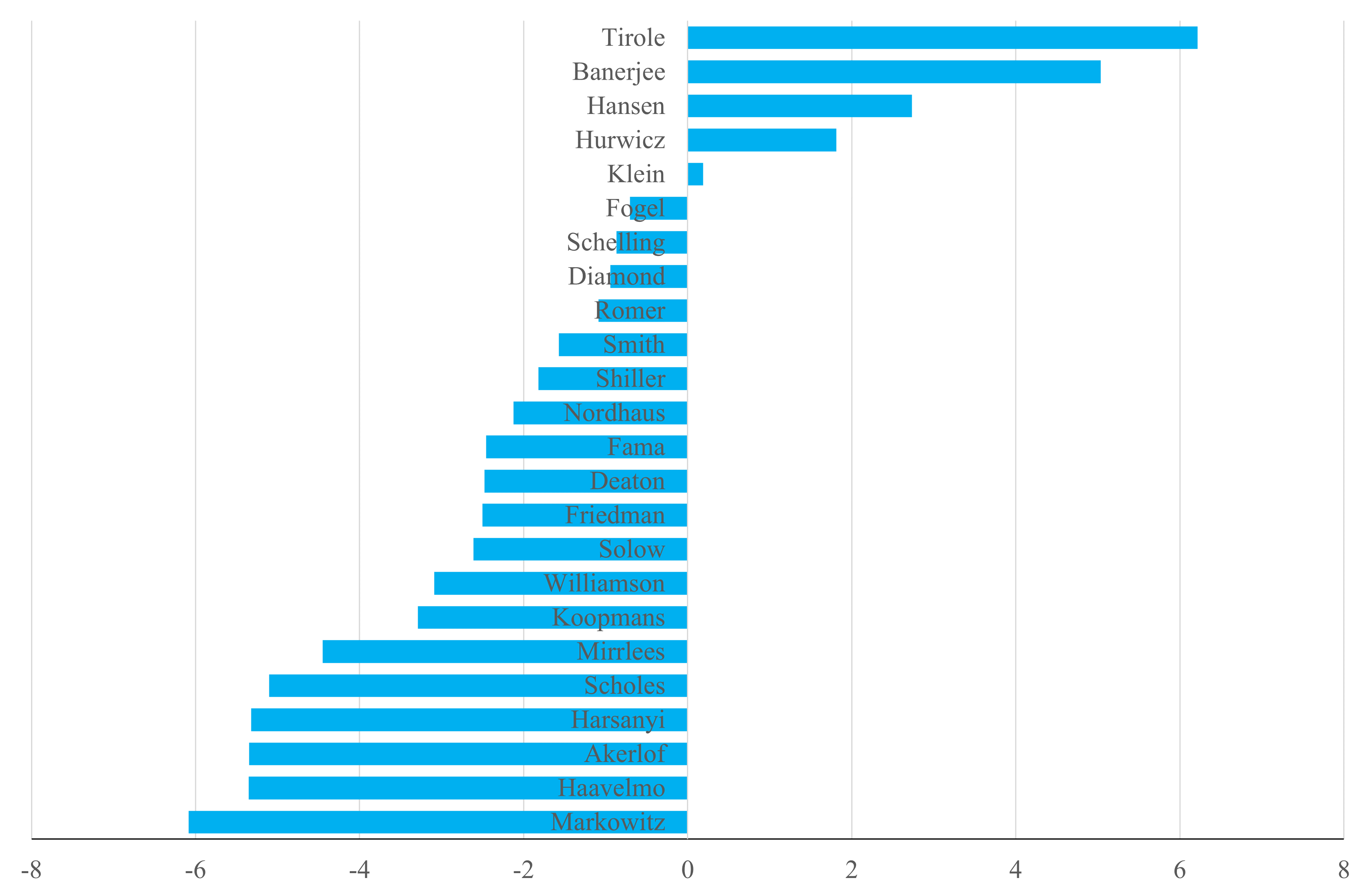}
    \caption{The t-statistic of the impact of a professor winning the Nobel prize on the annual citation numbers of his student who won the Nobel prize after him.}
    \label{fig:prof}
\end{figure}

\begin{figure}
    \centering
    \includegraphics[width=\textwidth]{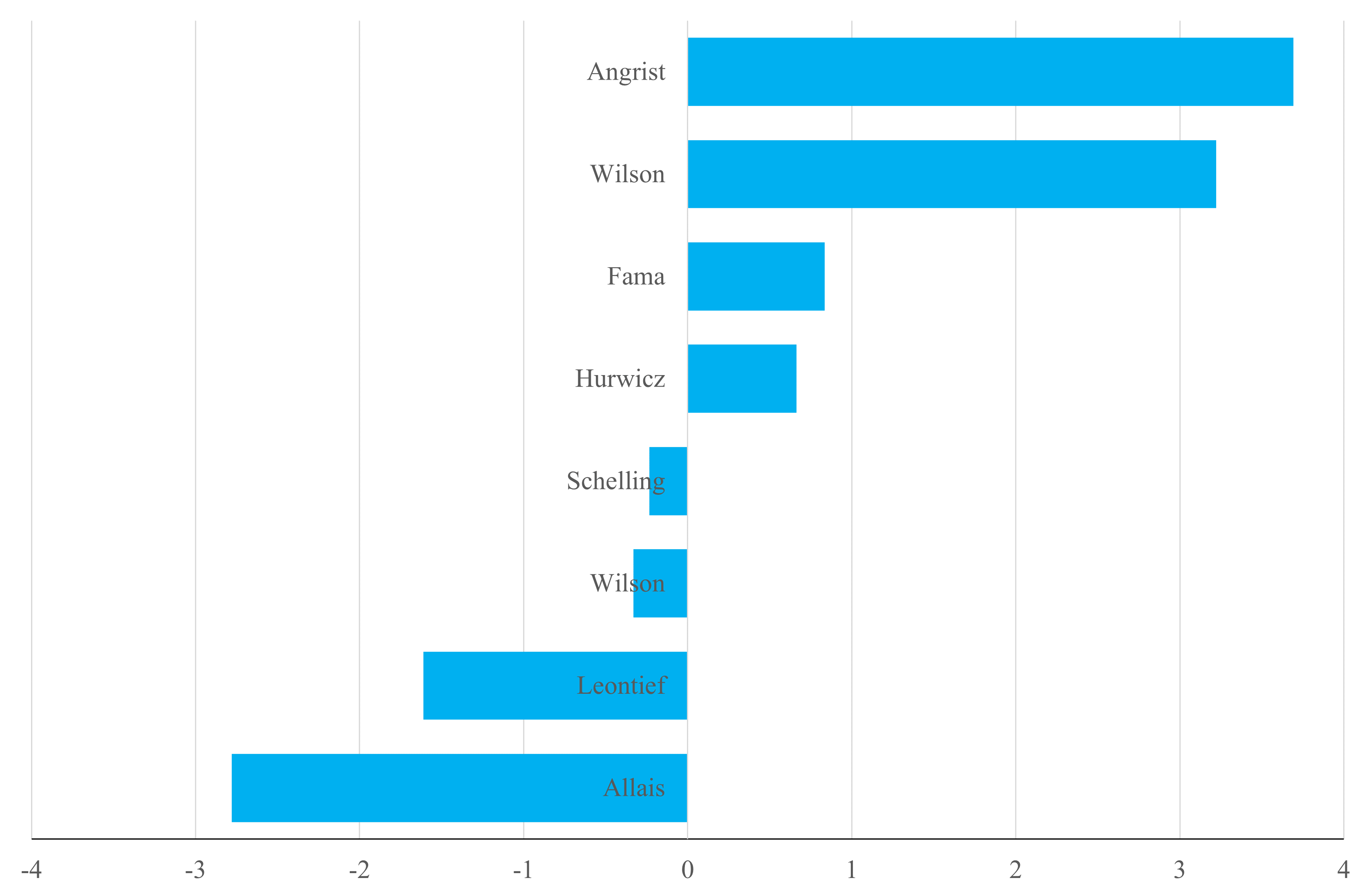}
    \caption{The t-statistic of the impact of a student winning the Nobel prize on the annual citation numbers of his or her professor who won the Nobel prize after him or her.}
    \label{fig:stud}
\end{figure}

\begin{figure}
    \centering
    \includegraphics[width=\textwidth]{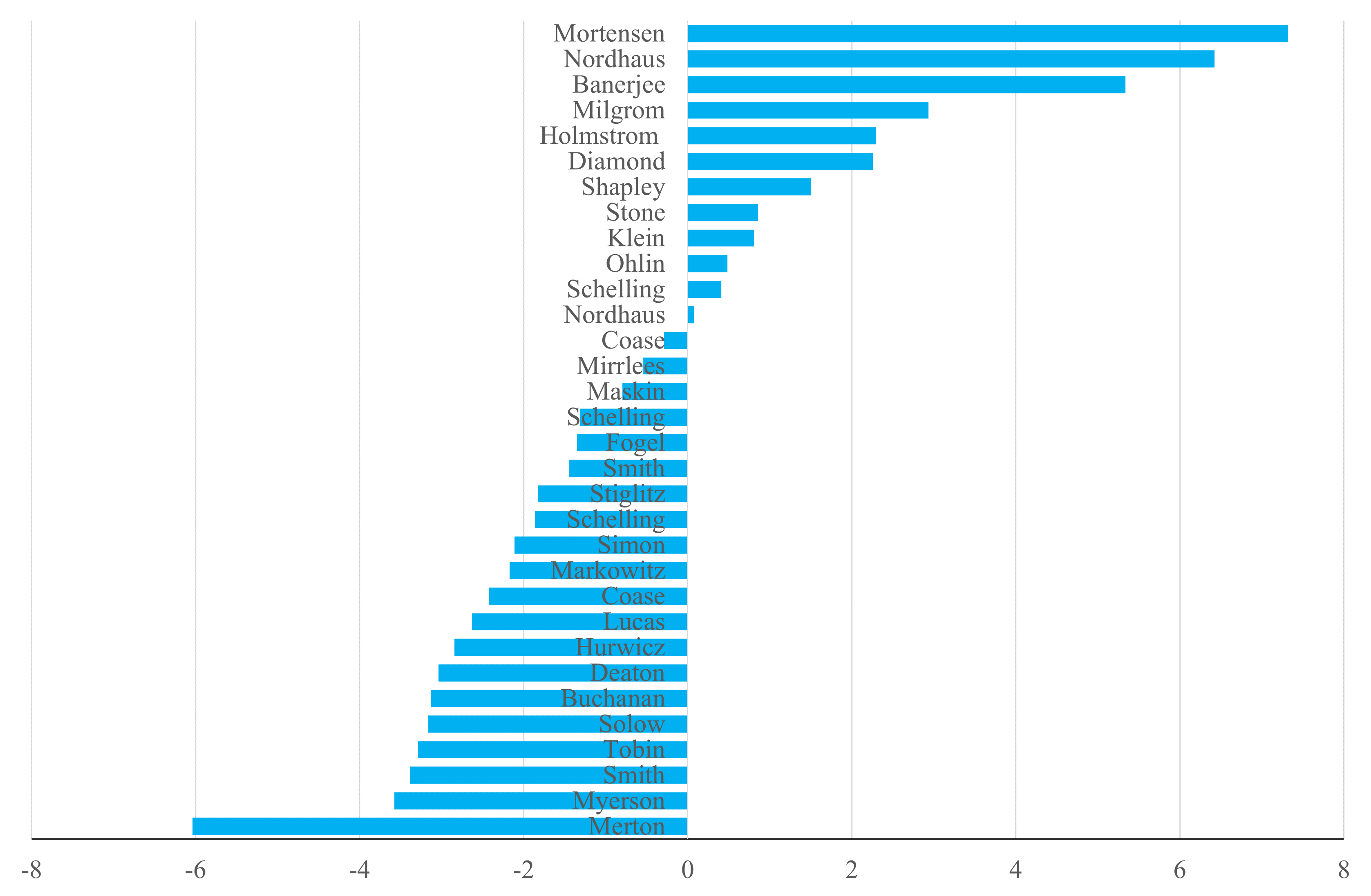}
    \caption{The t-statistic of the impact of a student winning the Nobel prize on the annual citation numbers of his fellow student who won the Nobel prize after him.}
    \label{fig:peer}
\end{figure}

\begin{figure}
    \centering
    \includegraphics[width=\textwidth]{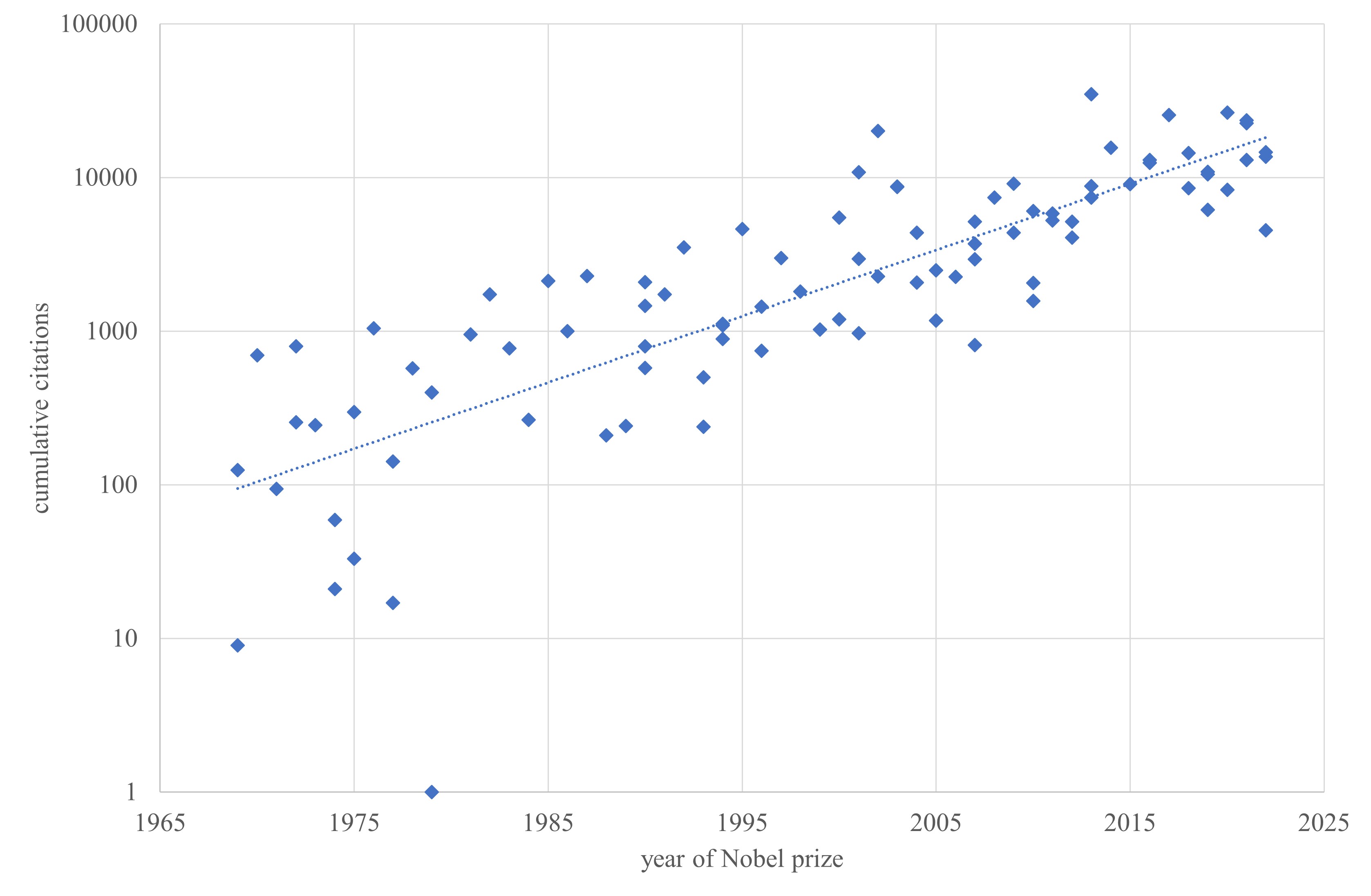}
    \caption{Cumulative citations (vertical axis) at the time of the Nobel award (horizontal axis) for all Laureates.}
    \label{fig:cumcit}
\end{figure}

\begin{longtable}{r l r r r l c c}
\caption{Nobel laureates and candidates.} \label{tab:names} \\
\hline ID & name & birth & death & won & alma mater & JEL & source \\ \hline 
\endfirsthead

\multicolumn{8}{c}%
{{\bfseries \tablename\ \thetable{} -- continued from previous page}} \\
\hline ID & name & birth & death & won & alma mater & JEL & source \\ \hline 
\endhead

\hline \hline
\endlastfoot

\hline \multicolumn{8}{r}{{Continued on next page}} \\ \hline
\endfoot
1	&	Ragnar Frisch 	&	1895	&	1973	&	1969	&	Oslo	&	E	&	-	\\
2	&	Jan Tinbergen 	&	1903	&	1994	&	1969	&	Leiden	&	E	&	-	\\
3	&	Paul Samuelson 	&	1915	&	2009	&	1970	&	Harvard	&	C	&	-	\\
4	&	Simon Kuznets 	&	1901	&	1985	&	1971	&	Columbia	&	O	&	-	\\
5	&	John Hicks 	&	1904	&	1989	&	1972	&	Oxford	&	C	&	-	\\
6	&	Kenneth Arrow 	&	1921	&	2017	&	1972	&	Columbia	&	D	&	-	\\
7	&	Wassily Leontief 	&	1905	&	1999	&	1973	&	Berlin	&	C	&	-	\\
8	&	Gunnar Myrdal 	&	1898	&	1987	&	1974	&	Stockholm	&	E	&	-	\\
9	&	Friedrich Hayek 	&	1899	&	1992	&	1974	&	Vienna	&	P	&	-	\\
10	&	Tjalling Koopmans 	&	1910	&	1985	&	1975	&	Leiden	&	C	&	-	\\
11	&	Leonid Kantorovich 	&	1912	&	1986	&	1975	&	Leningrad	&	C	&	-	\\
12	&	Milton Friedman 	&	1912	&	2006	&	1976	&	Columbia	&	E	&	-	\\
13	&	Bertil Ohlin 	&	1899	&	1979	&	1977	&	Stockholm	&	F	&	-	\\
14	&	James Meade 	&	1907	&	1995	&	1977	&	Cambridge	&	F	&	-	\\
15	&	Herbert Simon 	&	1916	&	2001	&	1978	&	Chicago	&	L	&	-	\\
16	&	Theodore Schultz 	&	1902	&	1998	&	1979	&	Wisconsin	&	O	&	-	\\
17	&	Arthur Lewis 	&	1915	&	1991	&	1979	&	LSE	&	O	&	-	\\
18	&	Lawrence Klein 	&	1920	&	2013	&	1980	&	MIT	&	C	&	-	\\
19	&	James Tobin 	&	1918	&	2002	&	1981	&	Harvard	&	G	&	-	\\
20	&	George Stigler 	&	1911	&	1991	&	1982	&	Chicago	&	D	&	-	\\
21	&	Gerard Debreu 	&	1921	&	2004	&	1983	&	Paris	&	D	&	-	\\
22	&	Richard Stone 	&	1913	&	1991	&	1984	&	Cambridge	&	E	&	-	\\
23	&	Franco Modigliani 	&	1918	&	2003	&	1985	&	New School	&	G	&	-	\\
24	&	James Buchanan 	&	1919	&	2013	&	1986	&	Chicago	&	H	&	-	\\
25	&	Robert Solow 	&	1924	&		&	1987	&	Harvard	&	O	&	-	\\
26	&	Maurice Allais 	&	1911	&	2010	&	1988	&	Paris	&	D	&	-	\\
27	&	Trygve Haavelmo 	&	1911	&	1999	&	1989	&	Oslo	&	C	&	-	\\
28	&	Merton Miller 	&	1923	&	2000	&	1990	&	Johns Hopkins	&	G	&	-	\\
29	&	Harry Markowitz 	&	1927	&		&	1990	&	Chicago	&	G	&	-	\\
30	&	William Sharpe 	&	1934	&		&	1990	&	Los Angeles	&	G	&	-	\\
31	&	Ronald Coase 	&	1910	&	2013	&	1991	&	LSE	&	K	&	-	\\
32	&	Gary Becker 	&	1930	&	2014	&	1992	&	Chicago	&	D	&	-	\\
33	&	Douglas North 	&	1920	&	2015	&	1993	&	Berkeley	&	N	&	-	\\
34	&	Robert Fogel 	&	1926	&	2013	&	1993	&	Johns Hopkins	&	N	&	-	\\
35	&	John Harsanyi 	&	1920	&	2000	&	1994	&	Stanford	&	C	&	-	\\
36	&	John Nash 	&	1928	&	2015	&	1994	&	Princeton	&	C	&	-	\\
37	&	Reinhard Selten 	&	1930	&	2016	&	1994	&	Frankfurt	&	C	&	-	\\
38	&	Robert Lucas 	&	1937	&		&	1995	&	Chicago	&	E	&	-	\\
39	&	William Vickrey 	&	1914	&	1996	&	1996	&	Columbia	&	D	&	-	\\
40	&	James Mirrlees 	&	1936	&	2018	&	1996	&	Cambridge	&	D	&	-	\\
41	&	Myron Scholes 	&	1941	&		&	1997	&	Chicago	&	G	&	-	\\
42	&	Robert Merton 	&	1944	&		&	1997	&	MIT	&	G	&	-	\\
43	&	Amartya Sen 	&	1933	&		&	1998	&	Cambridge	&	D	&	-	\\
44	&	Robert Mundell 	&	1932	&	2021	&	1999	&	MIT	&	E	&	-	\\
45	&	Daniel McFadden 	&	1937	&		&	2000	&	Minnesota	&	C	&	-	\\
46	&	James Heckman 	&	1944	&		&	2000	&	Princeton	&	C	&	-	\\
47	&	George Akerlof 	&	1940	&		&	2001	&	MIT	&	D	&	-	\\
48	&	Michael Spence 	&	1943	&		&	2001	&	Harvard	&	D	&	-	\\
49	&	Joseph Stiglitz 	&	1943	&		&	2001	&	MIT	&	D	&	-	\\
50	&	Vernon Smith 	&	1927	&		&	2002	&	Harvard	&	D	&	-	\\
51	&	Daniel Kahneman 	&	1934	&		&	2002	&	Berkeley	&	D	&	-	\\
52	&	Clive Granger 	&	1934	&	2009	&	2003	&	Nottingham	&	C	&	-	\\
53	&	Robert Engle 	&	1942	&		&	2003	&	Cornell	&	C	&	-	\\
54	&	Edward Prescott 	&	1940	&		&	2004	&	Carnegie Mellon	&	E	&	-	\\
55	&	Finn Kydland 	&	1943	&		&	2004	&	Carnegie Mellon	&	E	&	-	\\
56	&	Thomas Schelling 	&	1921	&	2016	&	2005	&	Harvard	&	C	&	-	\\
57	&	Robert Aumann 	&	1930	&		&	2005	&	MIT	&	C	&	-	\\
58	&	Edmund Phelps 	&	1933	&		&	2006	&	Yale	&	E	&	-	\\
59	&	Leonid Hurwicz 	&	1917	&	2008	&	2007	&	LSE	&	D	&	-	\\
60	&	Eric Maskin 	&	1950	&		&	2007	&	Harvard	&	D	&	-	\\
61	&	Roger Myerson 	&	1951	&		&	2007	&	Harvard	&	D	&	-	\\
62	&	Paul Krugman 	&	1953	&		&	2008	&	MIT	&	F	&	-	\\
63	&	Oliver Williamson 	&	1932	&	2020	&	2009	&	Carnegie Mellon	&	H	&	-	\\
64	&	Elinor Ostrom 	&	1933	&	2012	&	2009	&	Los Angeles	&	Q	&	-	\\
65	&	Dale Mortensen 	&	1939	&	2014	&	2010	&	Carnegie Mellon	&	D	&	-	\\
66	&	Peter Diamond 	&	1940	&		&	2010	&	MIT	&	D	&	-	\\
67	&	Christopher Pissarides 	&	1948	&		&	2010	&	LSE	&	D	&	-	\\
68	&	Christopher Sims 	&	1942	&		&	2011	&	Harvard	&	M	&	-	\\
69	&	Thomas Sargent 	&	1943	&		&	2011	&	Harvard	&	M	&	-	\\
70	&	Lloyd Shapley 	&	1923	&	2016	&	2012	&	Princeton	&	C	&	-	\\
71	&	Alvin Roth 	&	1951	&		&	2012	&	Stanford	&	D	&	-	\\
72	&	Eugene Fama 	&	1939	&		&	2013	&	Chicago	&	G	&	-	\\
73	&	Robert Shiller 	&	1946	&		&	2013	&	MIT	&	F	&	-	\\
74	&	Lars Peter Hansen 	&	1952	&		&	2013	&	Minnesota	&	F	&	-	\\
75	&	Jean Tirole 	&	1953	&		&	2014	&	MIT	&	L	&	-	\\
76	&	Angus Deaton 	&	1945	&		&	2015	&	Cambridge	&	I	&	-	\\
77	&	Oliver Hart 	&	1948	&		&	2016	&	Princeton	&	D	&	-	\\
78	&	Bengt Holmstrom 	&	1949	&		&	2016	&	Stanford	&	D	&	-	\\
79	&	Richard Thaler 	&	1945	&		&	2017	&	Rochester	&	D	&	-	\\
80	&	William Nordhaus 	&	1941	&		&	2018	&	MIT	&	Q	&	-	\\
81	&	Paul Romer 	&	1955	&		&	2018	&	Chicago	&	O	&	-	\\
82	&	Abhijit Banerjee 	&	1961	&		&	2019	&	Harvard	&	O	&	-	\\
83	&	Michael Kremer 	&	1964	&		&	2019	&	Harvard	&	O	&	-	\\
84	&	Esther Duflo 	&	1972	&		&	2019	&	MIT	&	O	&	-	\\
85	&	Robert Wilson 	&	1937	&		&	2020	&	Harvard	&	D	&	-	\\
86	&	Paul Milgrom 	&	1948	&		&	2020	&	Stanford	&	D	&	-	\\
87	&	David Card 	&	1956	&		&	2021	&	Princeton	&	J	&	-	\\
88	&	Joshua Angrist 	&	1960	&		&	2021	&	Princeton	&	C	&	-	\\
89	&	Guido Imbens 	&	1963	&		&	2021	&	Brown	&	C	&	-	\\
183 & Ben Bernanke & 1953 & & 2022 & MIT & E & \\
151	&	Douglas Diamond	&	1953	&		&	2022	&	Yale	&	G	&	\\
184	&	Philip Dybvig	&	1955	&		&	2022	&	Yale	&	G	&	\\
90	&	Ludwig von Mises	&	1881	&	1973	&		&	Vienna	&	P	&	ad hoc	\\
91	&	Frank Knight	&	1882	&	1972	&		&	Cornell	&	D	&	ad hoc	\\
92	&	Alvin Hansen	&	1887	&	1975	&		&	Wisconsin	&	E	&	ad hoc	\\
93	&	Harold Hotelling	&	1895	&	1973	&		&	Princeton	&	C	&	ad hoc	\\
185 & Roy Harrod & 1900 & 1978 & & Cambridge & E & ad hoc \\
94	&	Oskar Morgenstern	&	1902	&	1977	&		&	Vienna	&	D	&	ad hoc	\\
95	&	Abba Lerner	&	1903	&	1982	&		&	LSE	&	D	&	ad hoc	\\
96	&	Joan Robinson	&	1903	&	1983	&		&	Cambridge	&	E	&	ad hoc	\\
97	&	Ester Boserup	&	1910	&	1999	&		&	Copenhagen	&	O	&	ad hoc	\\
98	&	Edith Penrose	&	1914	&	1996	&		&	Johns Hopkins	&	M	&	ad hoc	\\
99	&	Lionel McKenzie	&	1919	&	2010	&		&	Princeton	&	D	&	ad hoc	\\
100	&	William Baumol	&	1922	&	2017	&		&	LSE	&	J	&	Clarivate	\\
101	&	Gordon Tullock	&	1922	&	2014	&		&	Chicago	&	K	&	Clarivate	\\
102	&	William Meckling	&	1922	&	1998	&		&	Chicago	&	G	&	ad hoc	\\
103	&	Jacob Mincer	&	1922	&	2006	&		&	Columbia	&	J	&	ad hoc	\\
186 & Kelvin Lancaster & 1924 & 1999 & & London & D & ad hoc \\
104	&	Henri Theil	&	1924	&	2000	&		&	Utrecht	&	C	&	ad hoc	\\
190 & Richard Easterlin & 1926 & & & Pennsylvania & O & Clarivate \\
105	&	Harold Demsetz	&	1930	&	2019	&		&	Northwestern	&	K	&	Clarivate	\\
106	&	Israel Kirzner	&	1930	&		&		&	New York	&	L	&	Clarivate	\\
107	&	Wayne Fuller	&	1931	&		&		&	Iowa	&	C	&	Clarivate	\\
191 & Richard Layard & 1934 & & & LSE & J & Clarivate \\
187 & Mancur Olson & 1932 & 1998 & & Maryland & D & ad hoc \\
108	&	Dale Jorgenson	&	1933	& 2022 &		&	Harvard	&	E	&	Clarivate	\\
109	&	Jagdish Bhagwati	&	1934	&		&		&	MIT	&	F	&	Clarivate	\\
110	&	Anne Krueger	&	1934	&		&		&	Wisconsin	&	H	&	Clarivate	\\
111	&	Amos Tversky	&	1937	&	1996	&		&	Michigan	&	D	&	ad hoc	\\
112	&	Fischer Black	&	1938	&	1995	&		&	Harvard	&	G	&	ad hoc	\\
193 & Samuel Bowles & 1939 & & & Harvard & D & Clarivate \\
113	&	Martin Feldstein	&	1939	&	2019	&		&	Oxford	&	E	&	Clarivate	\\
114	&	Michael Jensen	&	1939	&		&		&	Chicago	&	G	&	Clarivate	\\
115	&	Soren Johansen	&	1939	&		&		&	Copenhagen	&	C	&	Clarivate	\\
116	&	Richard Posner	&	1939	&		&		&	Harvard	&	K	&	Clarivate	\\
194 & Herbert Gintis & 1940 & & & Harvard & D & Clarivate \\
117	&	Sam Peltzman	&	1940	&		&		&	Chicago	&	H	&	Clarivate	\\
118	&	Stewart Myers	&	1940	&		&		&	Stanford	&	G	&	Clarivate	\\
119	&	Guillermo Calvo	&	1941	&		&		&	Yale	&	E	&	ad hoc	\\
120	&	Evan Porteus	&	1942	&		&		&	Case	&	D	&	ad hoc	\\
121	&	Martin Weitzman	&	1942	&	2019	&		&	MIT	&	Q	&	Clarivate	\\
122	&	Robert Hall	&	1943	&		&		&	MIT	&	E	&	Clarivate	\\
123	&	Mark Granovetter	&	1943	&		&		&	Harvard	&	D	&	Clarivate	\\
124	&	Katarina Juselius	&	1943	&		&		&	Helsinki	&	C	&	Clarivate	\\
125	&	Robert Barro	&	1944	&		&		&	Harvard	&	E	&	Clarivate	\\
126	&	Avinash Dixit	&	1944	&		&		&	MIT	&	D	&	Clarivate	\\
127	&	David Hendry	&	1944	&		&		&	LSE	&	C	&	Clarivate	\\
128	&	Stephen Ross	&	1944	&	2017	&		&	Harvard	&	G	&	Clarivate	\\
129	&	Anthony Atkinson	&	1944	&	2017	&		&	Cambridge	&	D	&	Clarivate	\\
130	&	Brian Arthur	&	1945	&		&		&	Michigan	&	D	&	Clarivate	\\
131	&	David Dickey	&	1945	&		&		&	Iowa	&	C	&	Clarivate	\\
132	&	Jerry Hausman	&	1946	&		&		&	Oxford	&	C	&	Clarivate	\\
133	&	Elhanan Helpman	&	1946	&		&		&	Harvard	&	F	&	Clarivate	\\
134	&	Hashem Pesaran	&	1946	&		&		&	Cambridge	&	C	&	Clarivate	\\
135	&	John Taylor	&	1946	&		&		&	Stanford	&	E	&	Clarivate	\\
136	&	Claudia Goldin	&	1946	&		&		&	Chicago	&	J	&	Clarivate	\\
137	&	Francine Blau	&	1946	&		&		&	Harvard	&	J	&	ad hoc	\\
138	&	Joel Mokyr	&	1946	&		&		&	Yale	&	N	&	Clarivate	\\
139	&	Jean-Jacques Laffont	&	1947	&	2004	&		&	Harvard	&	L	&	ad hoc	\\
140	&	Edward Lazear	&	1948	&	2020	&		&	Harvard	&	J	&	Clarivate	\\
141	&	Olivier Blanchard	&	1948	&		&		&	MIT	&	E	&	Clarivate	\\
142	&	Peter Phillips	&	1948	&		&		&	LSE	&	C	&	Clarivate	\\
143	&	Charles Manski	&	1948	&		&		&	MIT	&	C	&	Clarivate	\\
144	&	David Teece	&	1948	&		&		&	Pennsylvania	&	L	&	Clarivate	\\
145	&	Ariel Pakes	&	1949	&		&		&	Harvard	&	C	&	Clarivate	\\
146	&	David Kreps	&	1950	&		&		&	Stanford	&	D	&	Clarivate	\\
147	&	Halbert White	&	1950	&	2012	&		&	MIT	&	C	&	Clarivate	\\
148	&	Ariel Rubinstein	&	1951	&		&		&	Jerusalem	&	C	&	Clarivate	\\
149	&	Mark Gertler	&	1951	&		&		&	Stanford	&	E	&	Clarivate	\\
150	&	Richard Blundell	&	1952	&		&		&	LSE	&	J	&	Clarivate	\\
192 & Andrew Oswald & 1953 & & & Oxford & J & Clarivate \\
152	&	Kenneth Rogoff	&	1953	&		&		&	MIT	&	G	&	Clarivate	\\
153	&	Sanford Grossman	&	1953	&		&		&	Chicago	&	G	&	ad hoc	\\
154	&	John Moore	&	1954	&		&		&	LSE	&	G	&	Clarivate	\\
155	&	Kenneth French	&	1954	&		&		&	Rochester	&	G	&	Clarivate	\\
156	&	David Audretsch	&	1954	&		&		&	Wisconsin	&	L	&	Clarivate	\\
157	&	Gene Grossman	&	1955	&		&		&	MIT	&	F	&	Clarivate	\\
158	&	Nobuhiro Kiyotaki	&	1955	&		&		&	Harvard	&	G	&	Clarivate	\\
159	&	George Loewenstein	&	1955	&		&		&	Yale	&	D	&	Clarivate	\\
160	&	Carmen Reinhart	&	1955	&		&		&	Columbia	&	F	&	Clarivate	\\
161	&	Philippe Aghion	&	1956	&		&		&	Harvard	&	O	&	Clarivate	\\
162	&	Ernst Fehr	&	1956	&		&		&	Vienna	&	D	&	Clarivate	\\
163	&	Manuel Arellano	&	1957	&		&		&	LSE	&	C	&	Clarivate	\\
164	&	Daniel Levinthal	&	1957	&		&		&	Stanford	&	M	&	Clarivate	\\
165	&	Alberto Alesina	&	1957	&	2020	&		&	Harvard	&	H	&	Clarivate	\\
166	&	Kevin Murphy	&	1958	&		&		&	Chicago	&	D	&	Clarivate	\\
167	&	James Levinsohn	&	1958	&		&		&	Princeton	&	F	&	Clarivate	\\
168	&	Tim Bollerslev	&	1958	&		&		&	San Diego	&	C	&	IDEAS/RePEc	\\
169	&	John Campbell	&	1958	&		&		&	Yale	&	G	&	IDEAS/RePEc	\\
170	&	Stephen Berry	&	1959	&		&		&	Wisconsin	&	C	&	Clarivate	\\
171	&	Pierre Perron	&	1959	&		&		&	Yale	&	C	&	Clarivate	\\
172	&	Colin Camerer	&	1959	&		&		&	Chicago	&	D	&	Clarivate	\\
173	&	Robert Vishny	&	1959	&		&		&	MIT	&	G	&	IDEAS/RePEc	\\
174	&	Alan Krueger	&	1960	&	2019	&		&	Harvard	&	J	&	Clarivate	\\
198 & James Robinson & 1960 & & & Yale & O & Clarivate \\ 
175	&	Jordi Gali	&	1961	&		&		&	MIT	&	E	&	Clarivate	\\
176	&	Andrei Shleifer	&	1961	&		&		&	MIT	&	G	&	IDEAS/RePEc	\\
177	&	Stephen Bond	&	1963	&		&		&	Oxford	&	C	&	Clarivate	\\
189 & Simon Johnson & 1963 & & & MIT & O & Clarivate \\
178	&	Raghuram Rajan	&	1963	&		&		&	MIT	&	E	&	Clarivate	\\
179	&	Matthew Rabin	&	1963	&		&		&	MIT	&	D	&	Clarivate	\\
180	&	Daron Acemoglu	&	1967	&		&		&	LSE	&	O	&	Clarivate	\\
181	&	Marc Melitz	&	1968	&		&		&	Michigan	&	F	&	Clarivate	\\
182	&	John List	&	1968	&		&		&	Wyoming	&	D	&	Clarivate	\\
\end{longtable}

\end{document}